\documentclass[a4paper,12pt]{article}
\usepackage{amsmath}
\usepackage{amssymb}
\usepackage{graphicx}

\newcommand{\figdir}{.}

\newcommand{\doublefigure}[7][!htb]{   % \doublefigure[position]{label}{figname1}{figwidth1}{figname2}{figwidth2}{caption}
  \begin{figure}[#1]
    \begin{tabular}{cc}
      \begin{minipage}{#4}
                \includegraphics[width=#4]{#3}
                \center{\capfont{a.}}
      \end{minipage}
      &
      \begin{minipage}{#6}
        \includegraphics[width=#6]{#5}
        \center{\capfont{b.}}
      \end{minipage}
    \end{tabular}
    \caption{\capfont{#7}\label{#2}}
  \end{figure}
}

\newcommand{\putfigextra}[6][!htb]{ % \putfigextra[position]{label}{width}{name}{caption}{extra text}
    \begin{figure}[#1]
        \center\includegraphics[width=#3]{#4}
        \caption{\capfont{#5}\label{#2}}#6
    \end{figure}
}
\newcommand{\putfig}[5][!htb]{      % \putfig[position]{label}{width}{name}{caption}
    \putfigextra[#1]{#2}{#3}{#4}{#5}{}
}

\title{
\vskip10pt Competition of Languages and their Hamming Distance}
\newcommand{\affstyle}{\footnotesize\it}
\author{\vspace{-20pt}Tiberiu Te\c sileanu$^1$ and Hildegard~Meyer-Ortmanns$^2$\\
\\
\affstyle~School of Engineering and Science, International University Bremen\\
\affstyle~P.O.Box 750561, D-28725 Bremen, Germany\\
\affstyle~$^1$  t.tesileanu@iu-bremen.de \affstyle~\;\;$^2$
h.ortmanns@iu-bremen.de}
\date{}

\newcommand{\capfont}[1]{ \footnotesize{#1}\setlength{\baselineskip}{14pt} }

\begin{document}

\maketitle

\begin{abstract}
\setlength{\baselineskip}{12pt} \noindent We consider the
spreading and competition of languages that are spoken by a
population of individuals. The individuals can change their mother
tongue during their lifespan, pass on their language to their
offspring and finally die. The languages are described by
bitstrings, their mutual difference is expressed in terms of their
Hamming distance. Language evolution is determined by mutation and
adaptation rates. In particular we consider the case where the
replacement of a language by another one is determined by their
mutual Hamming distance. As a function of the mutation rate we
find a sharp transition between a scenario with one dominant
language and fragmentation into many language clusters. The
transition is also reflected in the Hamming distance between the
two languages with the largest and second to largest number of
speakers. We also consider the case where the population is
localized on a square lattice and the interaction of individuals
is restricted to a certain geometrical domain. Here it is again
the Hamming distance that plays an essential role in the final
fate of a language of either surviving or being extinct.
\\
\\
\noindent{\bf Keywords:} evolution of languages, population dynamics, bitstring models
\end{abstract}

\section{Introduction}

Evolution of language attracts the attention of linguists,
neuro- and computer-scientists, as well as physicists. It concerns
basic questions as the origin of speech, the role of mimicry and
movement, fine and rapid motor control, but also language
development \cite{others}. Language development refers to the
individual level during childhood, but also to language evolution
of mankind on a global scale. The evolution of a single language
as a function of time may refer to a transformation of this
language, say from ancient Latin to modern Italian.

When we talk about language evolution or language development in
this paper, we address the spreading, competition, extinction and
dominance of languages as a function of mutation and adaptation
rates that characterize the individual speakers. Similarly to the
evolution of cultural traits for various cultural features to a
homogeneous or multicultural polarized state \cite{miguel}, one
may ask here how an initially multilingual population evolves: do
the languages of individuals converge to a common ``lingua
franca'' in the very end, or is a fragmentation into many
coexisting language clusters the final state?

From daily-life experience it is intuitively clear that opposing
tendencies are at work: the convenience of having one common
language and the definition of national identity via language
favor language dominance or fragmentation, respectively. However,
when it comes to predictions for one of those scenarios, specific
mathematical modelling is needed to understand the transition
between dominance and fragmentation as various parameters change,
and to predict more specific features of this transition, {\it
e.g.} whether it is sharp or smooth.

Evolutionary models for language development have only recently
been proposed. There are macroscopic models, formulated in terms
of differential (rate) equations for population rates (that speak
a certain language)\cite{nowak}. They are macroscopic in the sense
that the fate of individuals is completely ignored for language
competition and the population rates refer to average quantities.
Stauffer and Schulze \cite{stauff} performed microscopic
simulations of language populations. Individuals inherit their
mother tongue from their parents, with a certain amount of
mutation. They change (adapt, replace) their language during their
lifespan, pass on their language to offspring, again with a
certain amount of mutation, and finally die. The time evolution of
the population follows a Verhulst dynamics %\cite{wiki_verhulst}
until a stable population size is achieved. The individuals are
described by bitstrings which code for their language. There are
as many different languages possible in the population as
bitstring states, that is $2^n$ if $n$ is the length of the
string.

In this paper we extend the simulations of Schulze and Stauffer to
include the measurement of the Hamming distance between the
dominating languages. We also generalize their model to implement
the Hamming distance in the interaction of individuals. A language
replacement is more likely to happen between languages with small
Hamming distance than with large one. Moreover, we locate the
population on a network (so far on a square lattice, for
simplicity) to simulate the influence of neighborhood. The
neighborhood can refer to ``external'' (geometric) or ``internal''
(social) space. Individuals influence their various languages
within a certain neighborhood, the so-called interaction range.

In the simulations so far \cite{stauff}, all positions of bits
within a bitstring are on an equal footing; we do not attempt any
mapping from features of languages like grammar, syntax,
vocabulary, to particular sequences of bits. Therefore, although
languages differing by just one bit in their bitstring description
would probably more appropriately be called dialects, our model is
too rough for such distinctions. In principle, it would be
possible to reserve a certain part of the bit-sequence to
grammatical features, others to the vocabulary and so on; the
rules for changing the bits would then depend on the position in
the string. Such generalizations are left for the future.

The paper is organized as follows. In section \ref{sec:model} we present our model,
in section \ref{sec:sim} we summarize the results of the simulations
and give a final outlook in section \ref{sec:outlook}.

\section{The Model}
\label{sec:model}

In this paper we consider extensions of the model presented in
\cite{stauff}. The first one deals with a population with random
interactions, independent of the spatial location of individuals,
the second one with a population living on a square lattice with
free and periodic boundary conditions, and a finite interaction
range.

\subsection{Language population with random interactions}
\label{sec:poploc}

We consider a population of individuals described by bitstrings of
length $n=32$; as argued in \cite{stauff}, $n = 30$ is already considered
sufficient for language simulations. The position of a bit inside
the string is given no specific meaning apart from its use as
coordinate to specify a certain language.
All $2^n$ possible different realizations of a bitstring are called languages. Two
different bitstrings are thought to represent different languages,  even if they
differ just by one bit.

The number of individuals,
{\it i.e.} the population size, evolves according to a Verhulst
dynamics. At each time step, an individual can either die, or live
and give birth to one offspring with probability $\alpha$. Here we
have chosen $\alpha=1$ in all simulations. This means that an
individual has one offspring once it survives at time step $t$.
Death occurs due to competition for resources, so that the
probability for dying is chosen proportional to the population
size $N$ at time $t$; that is, $\beta\cdot N(t)$. The population
size effectively follows a dynamics determined by
\begin{equation}\label{eq:verhulst}
\frac{dN(t)}{dt}\;=\; \alpha\cdot (1-\beta N(t))\cdot N(t)\;-\;\beta N(t)^2
\end{equation}
The gain term (first term) on the r.h.s.\!\! of
Eq.\eqref{eq:verhulst} reflects the rule that at time $t$, only
that fraction of the population which does not die at that
time-step [that is, $(1-\beta N(t))$], can have an offspring, with
probability $\alpha$. The loss term (second term) is quadratic in
$N(t)$ to implement the competition for resources: death occurs
with a probability proportional to the current population size. A
stationary population size  $N_0$ is obtained for
%\begin{equation}\label{eq:stationary}
%0\;=\; N_0\cdot (\alpha - (1+\alpha)\cdot \beta N_0)\;,
%\end{equation}
%or,
$\alpha$ and $\beta$ chosen such that
%\begin{equation}\label{eq:popsize}
$N_0\;=\;\frac{\alpha}{(1+\alpha)\cdot \beta}$.
%\end{equation}
In the simulation, we chose $\alpha$ and $N_0$ as independent
parameters, and adjusted $\beta$ accordingly.

As offspring are born, they inherit their ``mother'' tongue from
their parent (parent in the singular, because, for simplicity, we
consider asexual reproduction within this model; sexual
reproduction would offer the interesting option for offspring to
grow up bilingually \cite{mira}.) With a probability $p$, the
language of a newly-born can undergo a mutation, in which one
randomly chosen bit is switched from the value inherited from the
parent.

So far the model is similar to genetic models used in a biological
context, {\it e.g.} in the context of aging, in which individuals are
born and die while their genes suffer from (deleterious)
mutations. In contrast to genes, however, languages can be
completely exchanged by their speakers during lifetime. It is not
unusual for speakers of a language minority to change their
language to that spoken by the majority of population. To model
this behavior, an individual who speaks a language shared by
a fraction $x$ of the population, can change its language with probability
$q\cdot (1-x)^2$ to the language of a randomly chosen individual.
The parameter $q$ is called the language
change parameter. (Note that the notation here for $q$ is
different from that in \cite{stauff}.)

Large populations (strictly speaking, large population
\emph{densities}) usually increase the social pressure. Since it
is often the social pressure that leads to a language switch, we
further multiply $q\cdot (1-x)^2$ by $N(t)/N_0$ so that the
pressure is smaller in the expansion period than in the stable
state. The power of $(1-x)$, here chosen as two, can be used to
tune the ``social pressure''. The larger the exponent the less
likely a switch in language of the minority
speakers. The effective exponent is actually three, because there
is also the possibility that the randomly chosen person whose
language is the new candidate for a switch, speaks the same
language as the original one.

We extended the simulations of Schulze and Stauffer to measure the
Hamming distance between languages. (The Hamming distance between
two bitstrings is defined as the number of bits at corresponding
positions inside the string, by which the two strings differ.) The
Hamming distance is a quantitative measure for how much two
languages differ. In this paper we are interested in the Hamming
distance between the languages with the largest and the second
largest number of speakers. Are these languages similar or very
different? Moreover, this measure allows to implement and model
language barriers of the type that an Italian is more likely to
learn French than Chinese, since the learning effort is
proportional to the difference. In an extended version of the
former model, that we consider below, an individual with language
A switches to language B of another individual, randomly chosen
from the population together with $m-1$ other individuals, but
distinguished by the fact that language B comes closest to A in
terms of the Hamming distance. In case that several candidates
come closest to $A$, because they have the same shortest distance
to $A$, we take the first one out of the selection process. For
$m=1$ the model reduces to the former one. Again the switch is not
deterministic, but happens with probability $q\cdot (1-x)^2\cdot
N(t)/N_0$.

To summarize: the parameters that govern language dynamics in
these models, are the number $n$ of bits in a string coding a
specific language, the birth rate $\alpha$ for offspring, the
asymptotic population size $N_0$ (implicitly determining the death
rate $\propto \beta$), the language mutation rate $p$, the
language change parameter $q$, and, in  the extended version, the
number $m$ of individuals that is analyzed with respect to their
similarities, before the individual decides about a switch.

So far, all $2^n$ different languages, described by the state space of bit strings, are treated on an equal footing.
Effective weights come only from their number of speakers, but no intrinsic fitness was assigned to languages within these models.

\subsection{Language population on a lattice}
\label{sec:poplat} In this model, the ``language population'', is
placed on a square lattice. This assignment allows the description
of interactions that depend on the distance. The distance may
either represent the distance in Euclidean space or in internal
space, where ``internal'' can be {\it e.g.} ``social''. The
distance in Euclidean space plays a role for modelling situations
in which the pressure on people for learning a foreign language is
small when they live in the center of a large country rather than
at the border with another country. Due to modern communication
channels, this distance has lost its meaning to a certain extent,
because an individual, exposed to a foreign country, can keep and
speak its mother tongue with friends via phone or internet. In
this case, the friends would represent the nearest neighbors on a
square lattice in internal space. Of course, the regular features
of a square lattice look unrealistic as compared to network
topologies like small-world, hierarchical, or scale-free networks,
but as we shall see below, we use the regular features only for
defining the interaction range. Out of this interaction range,
{\it randomly} selected sites are then chosen for language
comparison to decide which language is taken over.

For simplicity we suppress the aspect of population evolution:
there are no birth or death rates, and each lattice site is always
occupied. To simplify the wording, from now on we will no longer
distinguish between an individual or speaker who speaks a certain
language, and the language itself, since individuals are
exclusively specified by their language (at a given site.) At each
time step, each individual can change its language with
probability $q$, to a different one chosen from those within a
given interaction range. The interaction range is defined as a
rectangular region centered at the individual. The half-height
$\Delta i$ and the half-width $\Delta j$ of this region are
parameters of the simulation. Both periodic and non-periodic
boundary conditions are implemented. In order to choose the
language to which an individual switches, we use a rule similar to
the former model: the individual randomly selects $m$ languages
out of the interaction range, and then chooses the one that is
closest to its own in terms of Hamming distance. Due to the random
selection within the interaction range, we actually use the
regular features of the square lattice in a very weak form. The
selected candidates may be nearest, next-to-nearest neighbors or
the like. The finite size of the square lattice enters in case of
non-periodic boundary conditions via the interaction range that is
chopped to fit into the simulation area when it touches the
boundary. In case of periodic boundary conditions, lattice sites
on opposite boundaries are identified as usual, so that the
interaction range is embedded in a torus.

In addition to language switches, at each time-step individuals
mutate their language with probability $p$ by a one-bit mutation,
as in the previous model by Schulze and Stauffer. Due to the
localization of speakers, the simulations amount to an animation
that shows the state of the language lattice at each time step.
Mechanisms such as language dominance or cluster formation are
easily visualized this way. (For language simulations on lattices
see also \cite{patri}.)

\section{Results}
\label{sec:sim}
\subsection{Language population with random interactions}

We performed simulations for population sizes $N_0$  ranging from
1000 to 100 000 (in factors of 10), for mutation rate and language
change parameters from $0$ to $1$ in various steps (see below),
and for lengths $n$ of the bitstrings with $16\leq n\leq 32$ in
steps of $1$. We considered cases with language switch independent
and dependent on the Hamming distance ($m=1$ and $m=4$,
respectively, in the notation of the previous section.) In order
to check the dependence on the initial conditions, we focused on
two cases; in the first case we studied one initial speaker with
language zero and let the population grow until it reached its
stable size; in the second case, the population was initially
comprised of the asymptotic number $N_0$ of individuals, half of
them speaking language $00_{(16)}$, and the other half speaking
$\mbox{FF}_{(16)}$. Here we are using a common notation for
hexadecimal numbers, that is $00_{(16)}$ $= 0\cdot 16^0 + 0\cdot
16^1 = 0$, $\mbox{FF}_{(16)}$ $= 15\cdot 16^0 + 15 \cdot 16^1 =
255$. In the bitstring, coding for language $255_{(10)}$ $\equiv
\mbox{FF}_{(16)}$, $n-8$ zero-bits are followed by $8$ one-bits,
where $n$ was usually chosen as $32$. We were interested in this
latter case to see whether competition of two equally represented
languages leads to the dominance of one of them or to the rise of
a third one.

Each simulation was run for $1000$ time-steps, and the following
parameters were recorded every $10$ steps
\begin{itemize}
\item
the number of individuals in the population $N(t)$
\item
the number of languages with more than ten speakers $n_l$
\item
the fraction of the population speaking the language of the majority $f_0(t)$
\item
the fraction of the population speaking the language with the second largest number of speakers $f_1(t)$
\item
the Hamming distance between the languages with the two largest numbers of speakers $d_0(t)$
\end{itemize}

After the last time-step of each simulation, we recorded these
parameters together with the bitstrings for the two ``largest''
languages in order to measure their Hamming distance. As it turned
out, already after less than a hundred iterations, $N(t)$, $n_1$,
$f_0(t)$, $f_1(t)$, and $d_0(t)$ reached their stable values in
both cases, $m=1$ and $m=4$, so that $1000$ time-steps were
sufficient for the population to stabilize in these properties.

\subsection{Language switch independent of the Hamming distance}

In our simulations, after $1000$ time-steps, we observed
essentially two scenarios. In the first one the language
population is characterized by {\it dominance} of one language
that comprises the largest part of the population, while the rest
of the population is scattered between many small languages. The
second scenario is {\it fragmentation}, a state in which no
language has significantly more speakers than the others. The
dominance of languages is accompanied by the fact that in this
phase the Hamming distance between the two largest languages is
usually equal to $1$, that is, the smallest possible value for two
different languages. Therefore, an even larger part of the
population speaks almost the same language. In the absence of a
dominant language, the Hamming distance between the two largest
languages is large, about $16$ (for $n = 32$) in our simulations.

The way we have implemented the switch to another language
supports language dominance, since speakers of a minority language
have a high probability to switch to the majority language when
they randomly select the language to which they switch. On the
other hand, mutation acts toward fragmentation by spreading the
speakers in language space. The result of these competing
tendencies can hardly be predicted intuitively. It was therefore
analyzed in the simulations. The transition between these two
cases is sharp, it looks like first order, and, all other
parameters fixed, occurs at a critical value of the mutation rate,
$p_0$. This can be seen, for instance, in Figure
\ref{fig:res:transition1},
%%%%%%%%
\doublefigure[!t]{fig:res:transition1}{\figdir/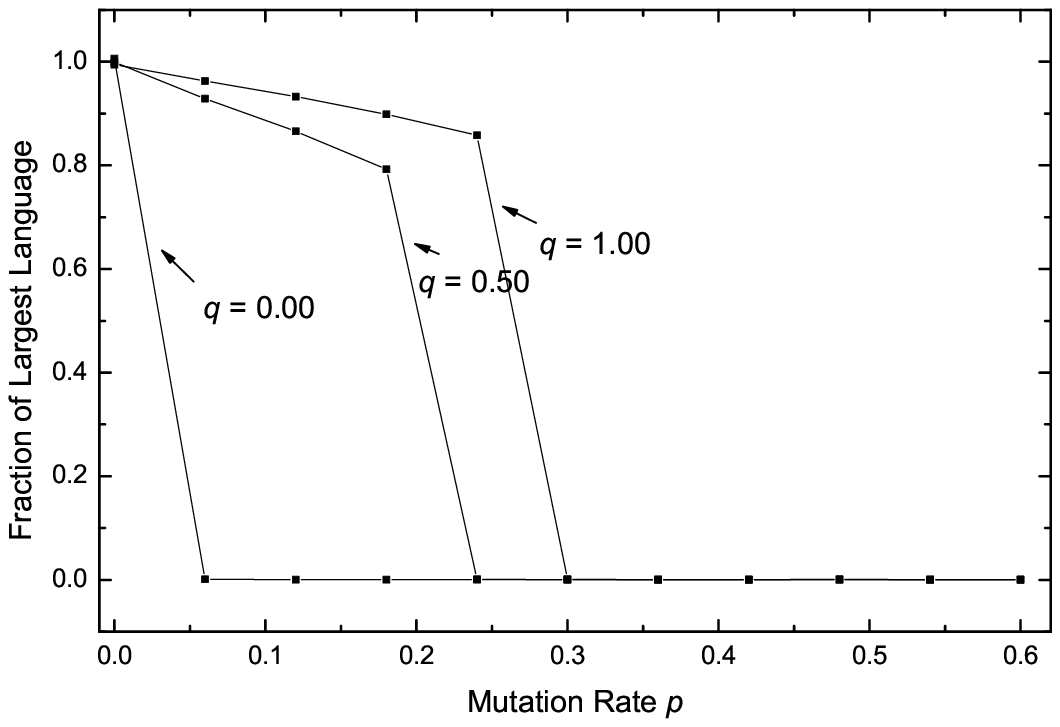}{7.5cm}{\figdir/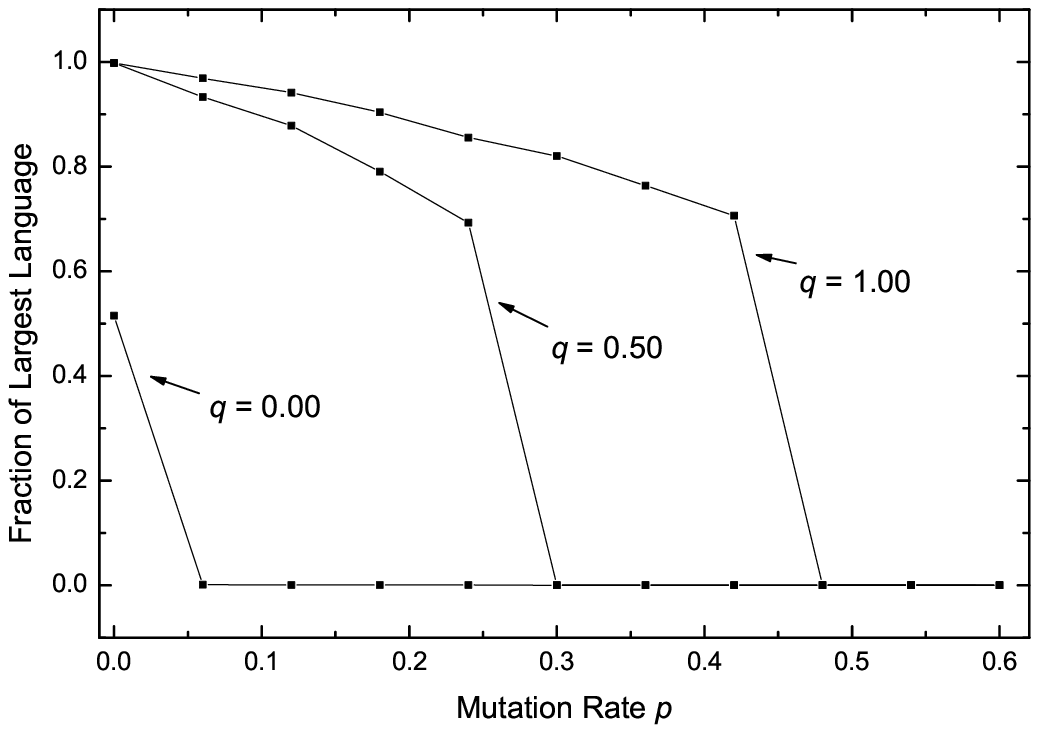}
{7.5cm}{Dependence of the relative size of the largest language on
the mutation rate $p$, for various language change parameters $q$.
$N_0 = $100 000, $n=32$. {\it a.} Initially just one speaker of
language zero. {\it b.} Initially $N_0$ individuals, speaking
languages $00_{(16)}$ and $\mbox{FF}_{(16)}$ (in hexadecimal)
equally.}
%%%%%%%%
where the relative size of the largest language is large for
$p<p_0$ and almost zero for $p>p_0$ (in Figure
\ref{fig:res:transition1}a the initial population contained only
one speaker of language zero, while it had $N_0$ speakers in
Figure \ref{fig:res:transition1}b, half speaking language
$00_{(16)}$, half $\mbox{FF}_{(16)}$. Therefore, initially, also
in the latter case only the eight most-right bits out of the $32$
are "used", i.e. they are one, but later, during the simulation,
all other $24$ bits may be switched to one due to the random
mutations.) As mentioned in \cite{stauff}, we also observed that
at values $p$ close to $p_0$, it may be possible to obtain both
dominance and fragmentation in different runs of the simulation.

The position $p_0$ of the phase transition depends on a number of
factors. Since the mutation rate $p$ acts toward fragmentation,
while the language change parameter  $q$ acts toward dominance,
$p_0$ will obviously depend on $q$ as it is already seen from \nolinebreak
\mbox{Figure \ref{fig:res:transition1}.} Figure \ref{fig:res:trans1q}
%%%
\putfig[t]{fig:res:trans1q}{10cm}{\figdir/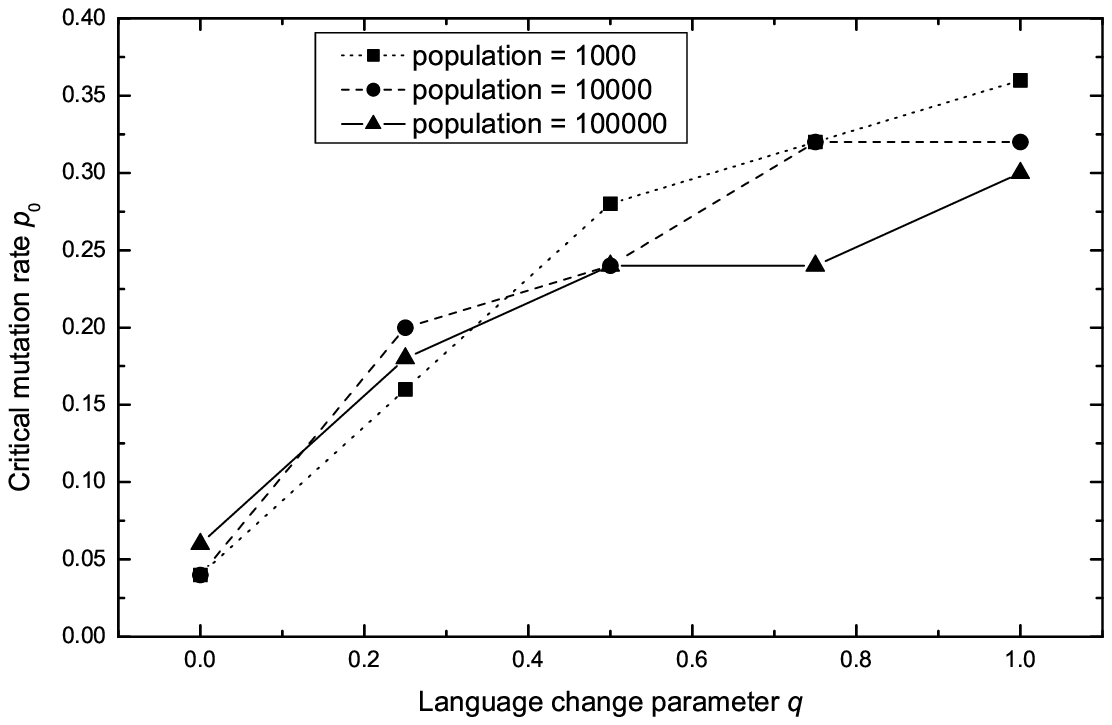}{Dependence
of the critical mutation rate $p_0$ on the language change
parameter $q$, at different population sizes; initially one
speaker of language zero, $n=32$.}
%%%
shows the dependence of the critical mutation rate on the language
change parameter $q$ for different population sizes. As observed
in \cite{stauff}, $p_0$ also depends on the size of the
population, although the tested range was varied only between
$1000$ and 100 000 individuals, while in \cite{stauff} it was
varied up to $10^7$.

%%%
\putfig{fig:res:trans2q}{10cm}{\figdir/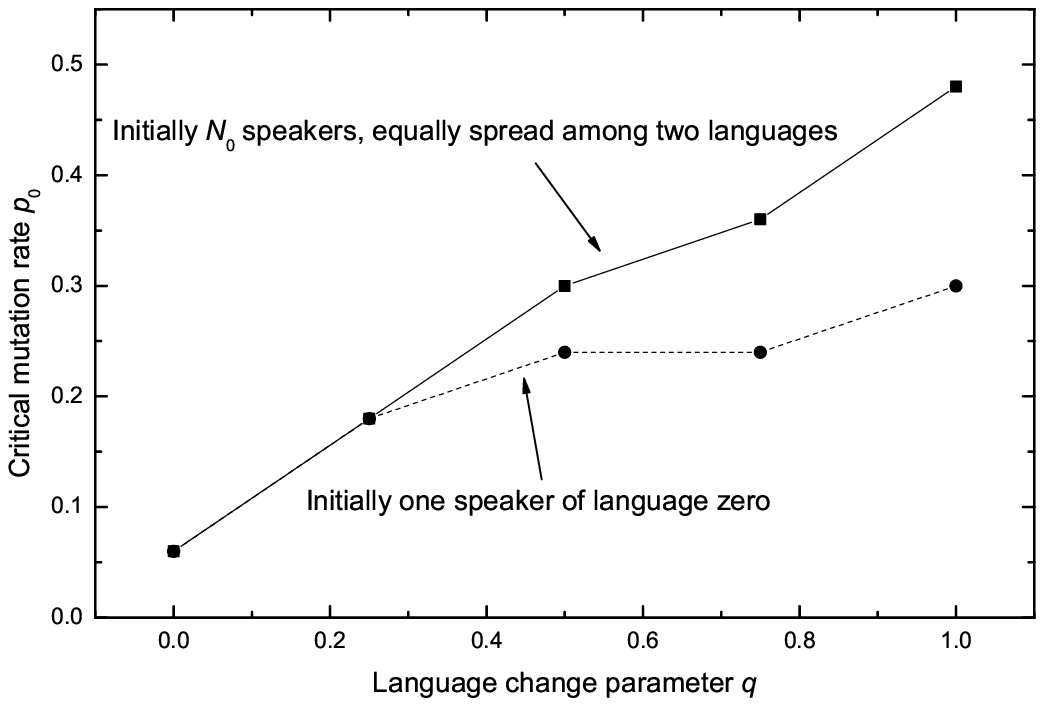}{Dependence
of the critical mutation rate $p_0$ on the language change
parameter $q$. $N_0 = $100 000, $n=32$.}
%%%%%%%%

The location of the phase transition also depends on the initial
conditions of the simulation. In case the initial population
starts with many speakers of two evenly represented languages,
here chosen as $00_{(16)}$ and $\mbox{FF}_{(16)}$, the critical
mutation rates are larger than in the case of initially one
individual, especially for large values of the language change
parameter $q$, Figure \ref{fig:res:trans2q}. This effect can be
explained by the fact that it is harder to break the initially big
domains of languages, in contrast to the single speaker  of
language zero in the other case.

As we have seen above, in the case of dominance, one language
has many more speakers than all the others. Due to mutation,
a fraction of these speakers switches at each time step to a
language that differs from the largest one by a single bit.
In the stationary state, the number of speakers who switch
to such a language, is approximately equal to the number of speakers
who switch back to the largest population due to the option of a
language switch. Since in this realization a single
mutation cannot change the language by more than one bit,
languages that differ by two or more bits from the largest one
are disfavored.

The effect is seen in measurements of the Hamming distance (number
of different bits) between the two largest languages. It is equal
to one (rarely two) before the transition $(p<p_0)$, and large
(around $16$ for $n=32$ bits per language) after the transition.
Figure \ref{fig:res:hambig}
%%%%
\doublefigure[t]{fig:res:hambig}{\figdir/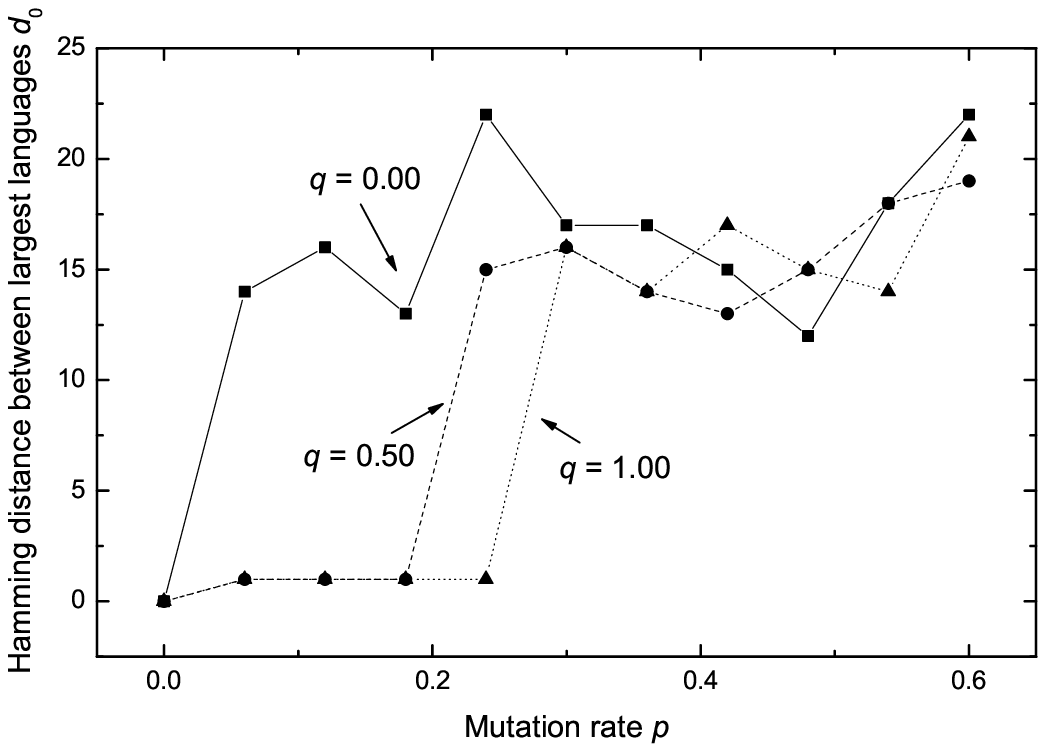}{7.5cm}{\figdir/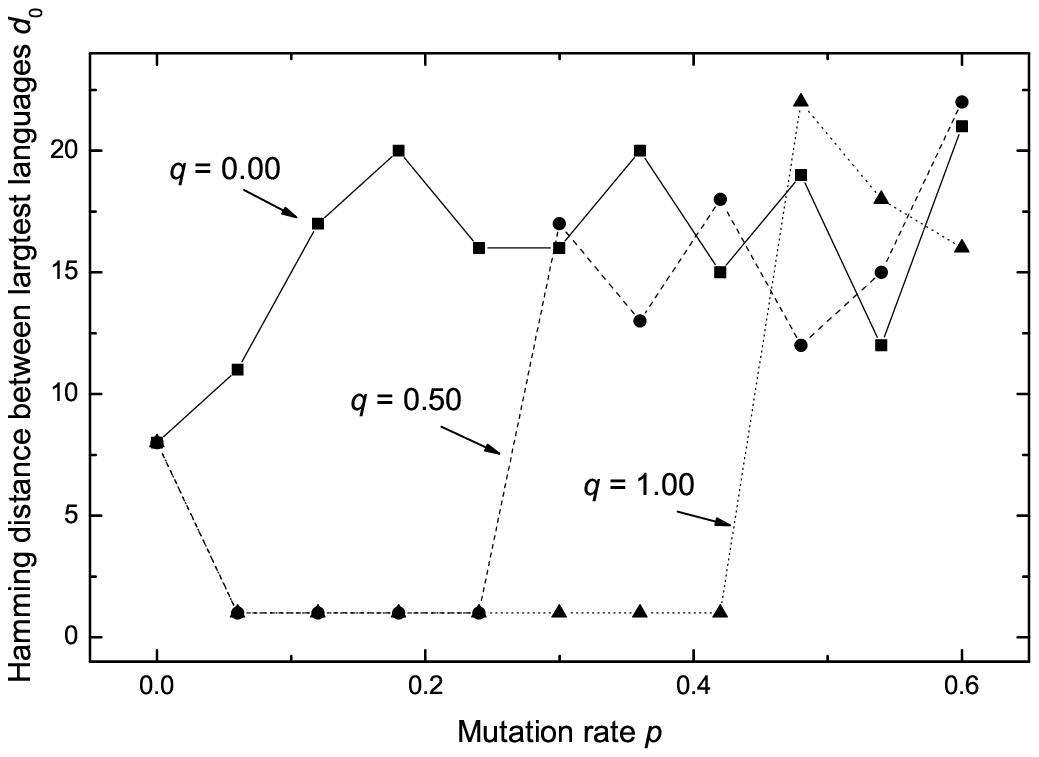}
{7.5cm}{Phase transition in the Hamming distance between the two
largest languages. The asymptotic population size is $N_0 =$100
000 in both cases, $n=32$. {\it a.} Initially one speaker of
language zero. {\it b.} Initially $N_0$ speakers evenly
distributed among languages $00_{(16)}$ and $\mbox{FF}_{(16)}$
(hexadecimal.)}
%%%%
shows this effect also in dependence on the initial conditions. The
coincidence of the transition in the Hamming distance with the one
in the relative size of the largest language can be seen in
Figure \ref{fig:res:hamsize}.
%%%%
\doublefigure[t]{fig:res:hamsize}{\figdir/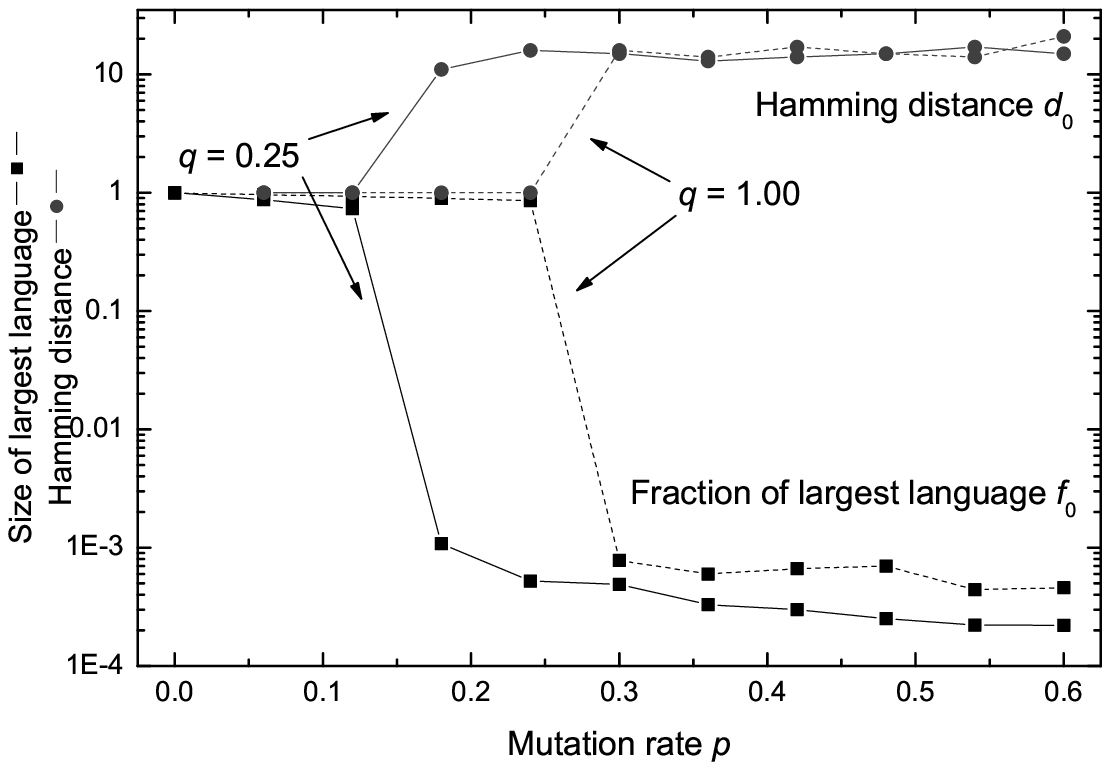}{7.5cm}{\figdir/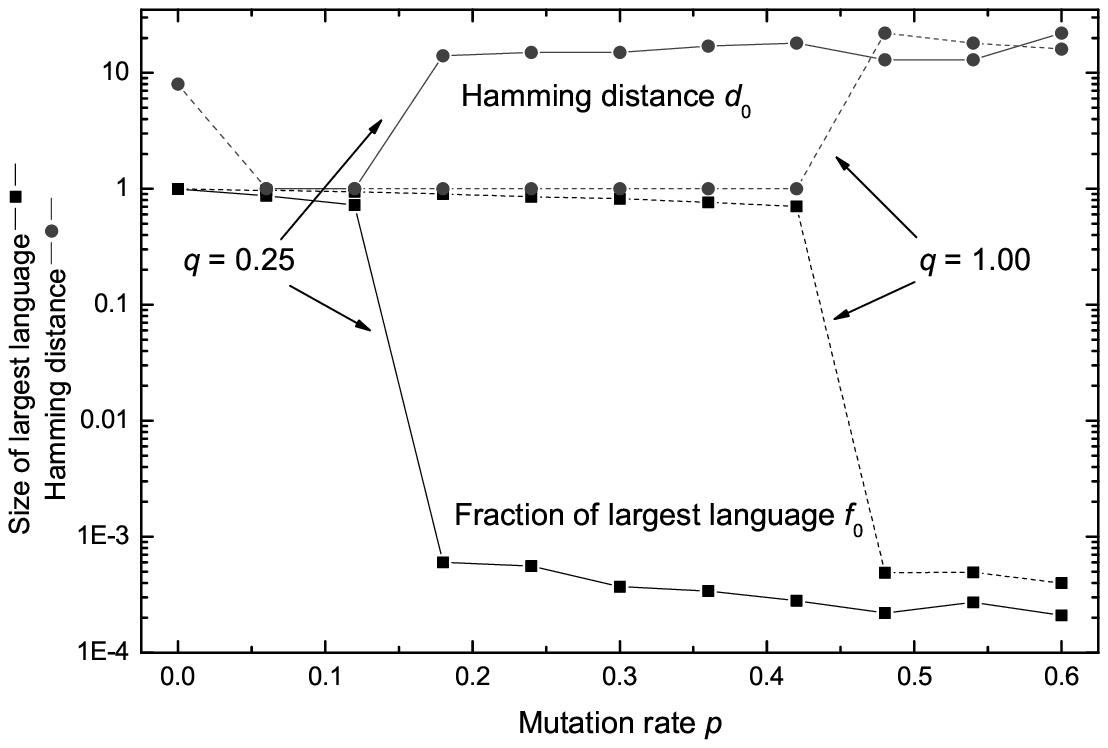}
{7.5cm}{Coincidence of phase transition in Hamming distance
between the two largest languages, with the one in the relative
size of the largest language. The asymptotic population size is
$N_0 = $100 000. The solid line corresponds to $q = 0.25$, and the
dashed one to $q = 1.00$. {\it a.} Initially one speaker of
language zero. {\it b.} Initially $N_0$ speakers evenly
distributed among languages $00_{(16)}$ and $\mbox{FF}_{(16)}$
(hexadecimal.)}
%%%%

Moreover, we observed that in case of dominance, the final largest
language is one of the initial languages, possibly with one or two
bits mutated. In the case where initially there is only one
speaker of language zero, the final dominant language is thus
expected to be language zero or a one- or two-bit mutation of it.
Initial conditions with languages $00_{(16)}$ and
$\mbox{FF}_{(16)}$ lead to a final dominant language that is one
of the initial two, up to a few mutations. Therefore we can
neither describe the evolution from ancient Latin to modern
Italian as long as one-bit mutations are unbiased, nor the
creation of English out of ancient French and ancient German
within this model, as long as we only allow one-bit mutations and
complete switches, but no ``superposition'' of languages in
analogy to the mixing of genes in biology (cf. \cite{stauff1} for
a proposal of a model in this direction.)

The previous results were obtained  by using $n=32$ bits per
language. The position of the phase transition is largely
independent of the bitlength $n$. This can be expected because the
language switch, which is opposing dominance, is itself
independent of $n$. The critical mutation rate for various
bitlengths is plotted in Figure \ref{fig:res:nbits}, for the case
where in the initial state there was one speaker of language zero.
We observed the same effect when we had in the initial state $N_0$
speakers equally spread among two languages.

In view of computational aspects we remark that bitstrings as long
as $32$ bits (providing a large language space, but for the case
of less large population size), were efficiently treated by using
a hash-table, so that the complexity of the simulation was $O(N)$,
i.e. linear in the population size.
%%%
\putfig[t]{fig:res:nbits}{10cm}{\figdir/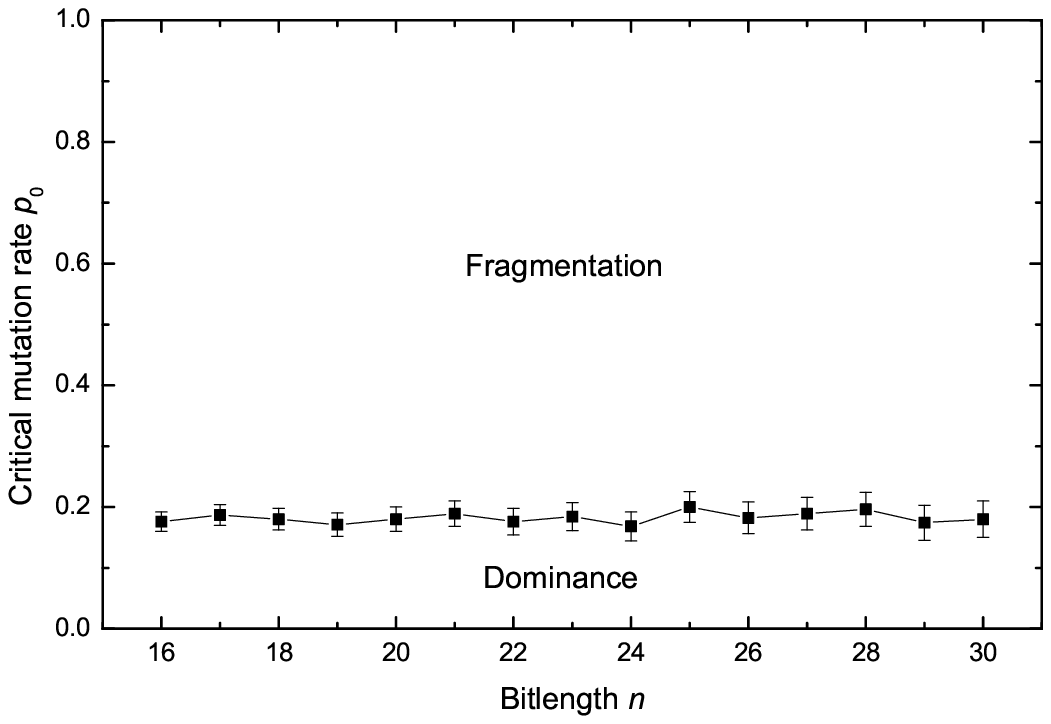}{Dependence
of the critical mutation rate on the bitlength $n$. Asymptotic
population size is $N_0 =$10 000, and initial state contained one
speaker of language zero. $q = 0.25$.}
%%%

\subsection{Language switch determined by the Hamming distance}

In case the language switch depends on the Hamming distance, as
described in section \ref{sec:poploc}, the formation of language clusters is
more favored than the dominance scenario as compared to the
previous model $(m=1)$, since each individual is more conservative
in changing its language. Therefore the sharp transition that was
observed for $m=1$ between dominance and fragmentation is smoothed
out, see Figure \ref{fig:res:notrans}.
%%%
\doublefigure{fig:res:notrans}{\figdir/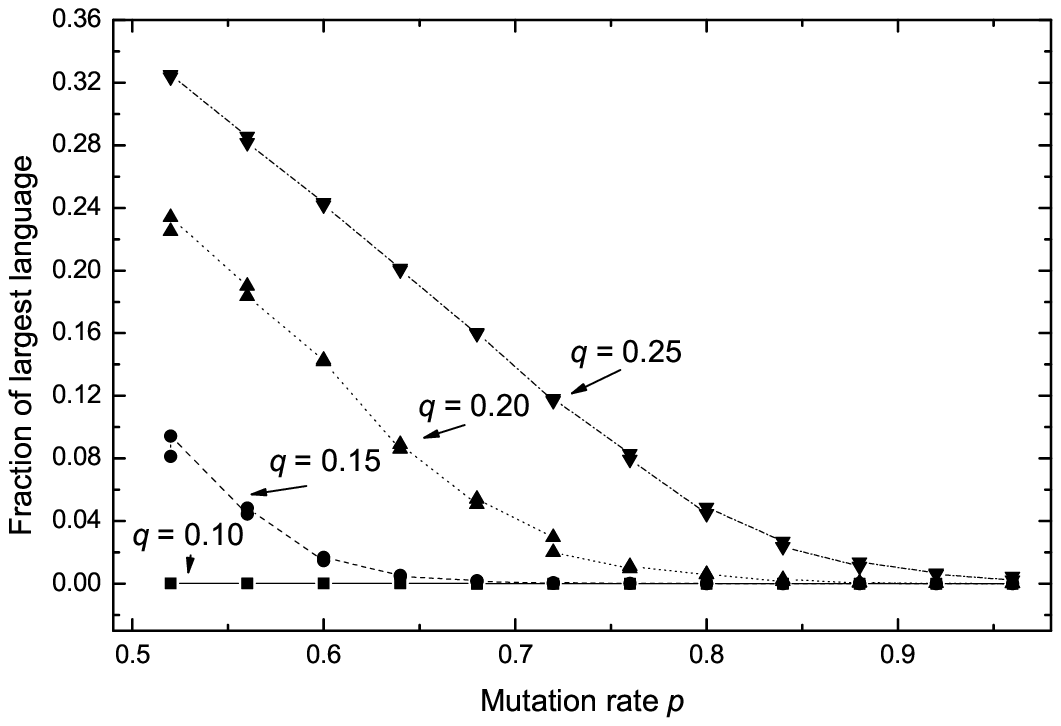}{7.5cm}{\figdir/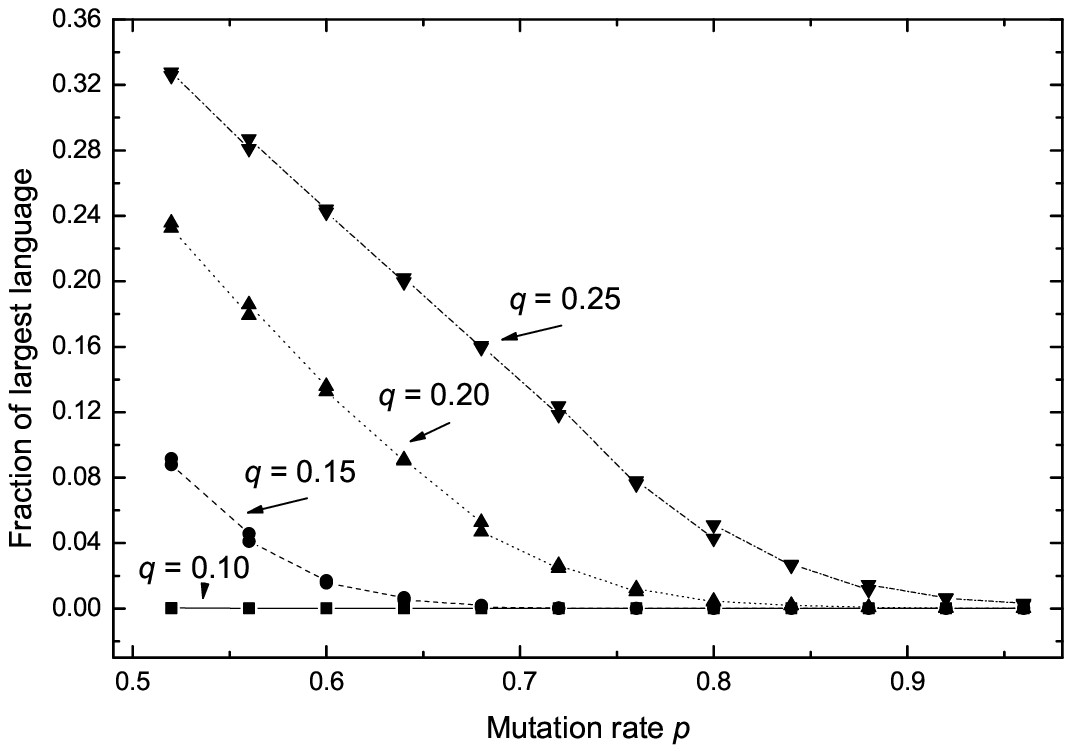}
{7.5cm}{Smooth crossover in case of Hamming-distance dependent
language change (for $m = 4$.) Asymptotic population size is $N_0
=$ 100 000 in both cases. {\it a.} Initially one speaker of
language zero. {\it b.} Initially $N_0$ speakers evenly
distributed among languages $00_{(16)}$ and $\mbox{FF}_{(16)}$
(hexadecimal.)}
%%%%
Moreover, the dependence on the initial conditions is less strong
than before, there is no noticeable difference between the
evolution for one or two initial languages, cf. Figures
\ref{fig:res:notrans}a and \ref{fig:res:notrans}b. This supports
the idea of the previous subsection that the dependence on the
initial conditions is related to the language dominance, and since
dominance in this model is disfavored, the influence due to
dominance should be less visible.

The crossover from dominance to fragmentation is still seen as a
sharp transition in the Hamming distance between the two largest
languages. As it is seen from Figure \ref{fig:res:notransham},
%%%%
\doublefigure[bt]{fig:res:notransham}{\figdir/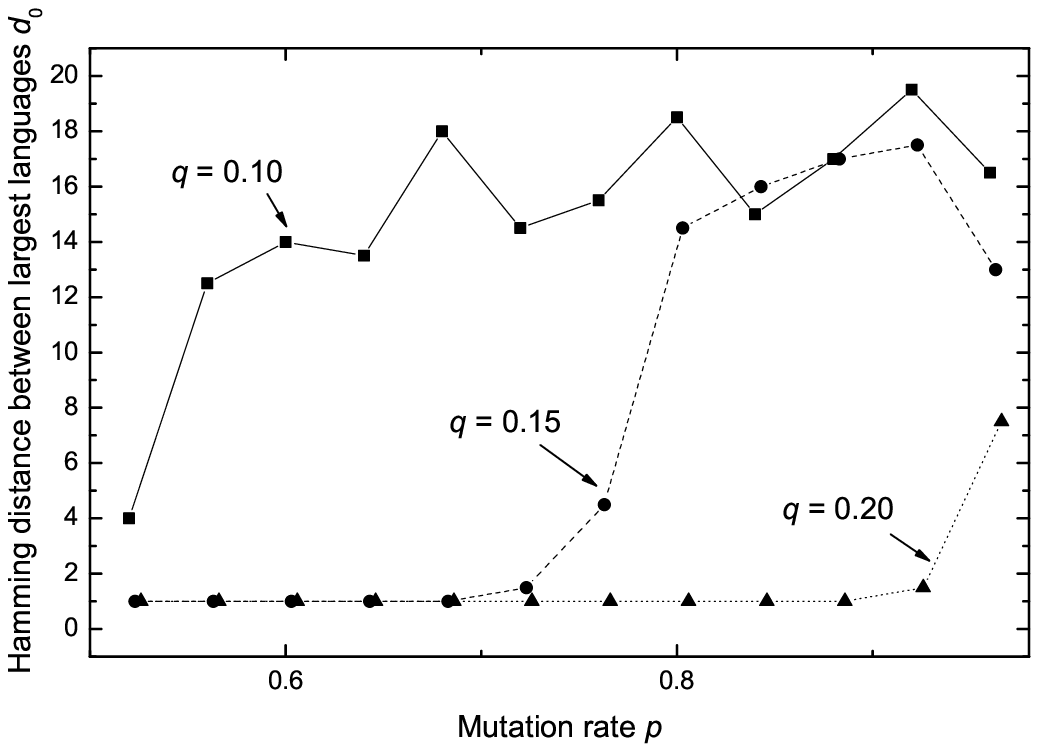}{7.5cm}
{\figdir/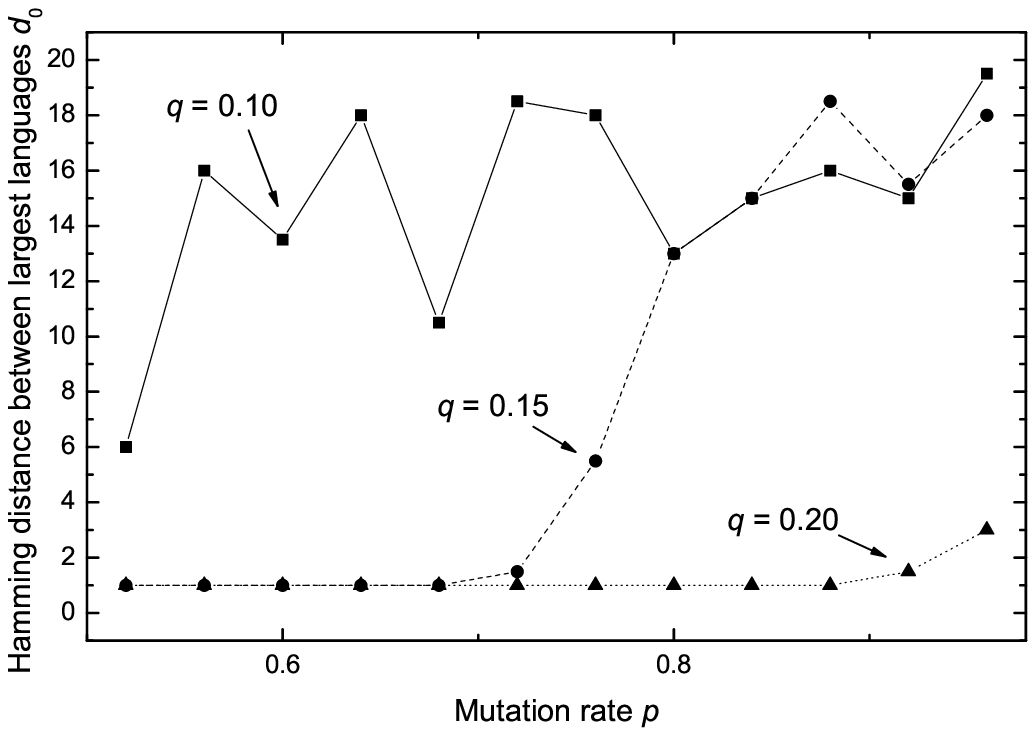}{7.5cm}{Phase transition in the Hamming
distance between the two largest languages, in the case of
Hamming-distance dependent language change (with $m = 4$.) Values
are averages of two runs for each $p$. Asymptotic population size
is $N_0 =$ 100 000 in both cases. {\it a.} Initially one speaker
of language zero. {\it b.} Initially $N_0$ speakers evenly
distributed among languages $00_{(16)}$ and $\mbox{FF}_{(16)}$.}
%%%%%%%%%%%%
the Hamming distance is equal to one up to a certain critical
mutation rate $p_0$ and then rises rather abruptly to values
larger than $10$. Again, the position of $p_0$ is relatively
independent on whether there were initially one or two languages.

We also observed a large variation of the Hamming distance for
different runs at the same values of the parameters; changes as high
as 30\% were observed in the case of fragmentation.

\subsection{Language population on a lattice}

In this paper we report on results of language evolution on a
lattice that allow a visualization of the results, that is, we
made snapshots of the language population on the square lattice
after each iteration. The simulations were performed on a square
lattice of size $400\times 400$ with free boundary conditions,
$n=16$ bits, mutation rate $p=0.01$, language change rate
$q=0.50$, and interaction range $41\times 41$. The number of
individuals considered for language change was $m=2$. As it turned
out in the simulations, the parameters $p$ and $q$ of language
mutation and language switch are of little relevance for the
outcome compared to the initial conditions, in contrast to the
simulations of the $m=1$ model. Because of the similarity with the
$m=4$ model of the previous section, it is not surprising that we
found a smooth crossover from the dominance scenario, in which
only a few languages survive, to the complete fragmentation. Note
that the $m=4$ model and the $m=2$ lattice population essentially
differ by the location of the interaction range, here centered at
an individual and extended to a finite (possibly small) subset of
the lattice in its neighborhood. The former simulation was
non-localized, {\it i.e.} any subset of $m$ speakers could be
selected from the whole population. This difference is probably
responsible for the stronger dependence on the initial conditions
that is observed here: when the mutation rate is chosen small
enough and the language change parameter large enough to form
stationary language clusters, the number of such clusters and
their location and shape are less likely ``washed out'' by random
interactions, but determined by the initial conditions of the
simulation.

Based on the description of the model of section \ref{sec:poplat}, a few
general features of these simulations can be deduced. It is seen
in Figure \ref{fig:res:whyflat}
%%%%%%%%%
\putfig{fig:res:whyflat}{10cm}{\figdir/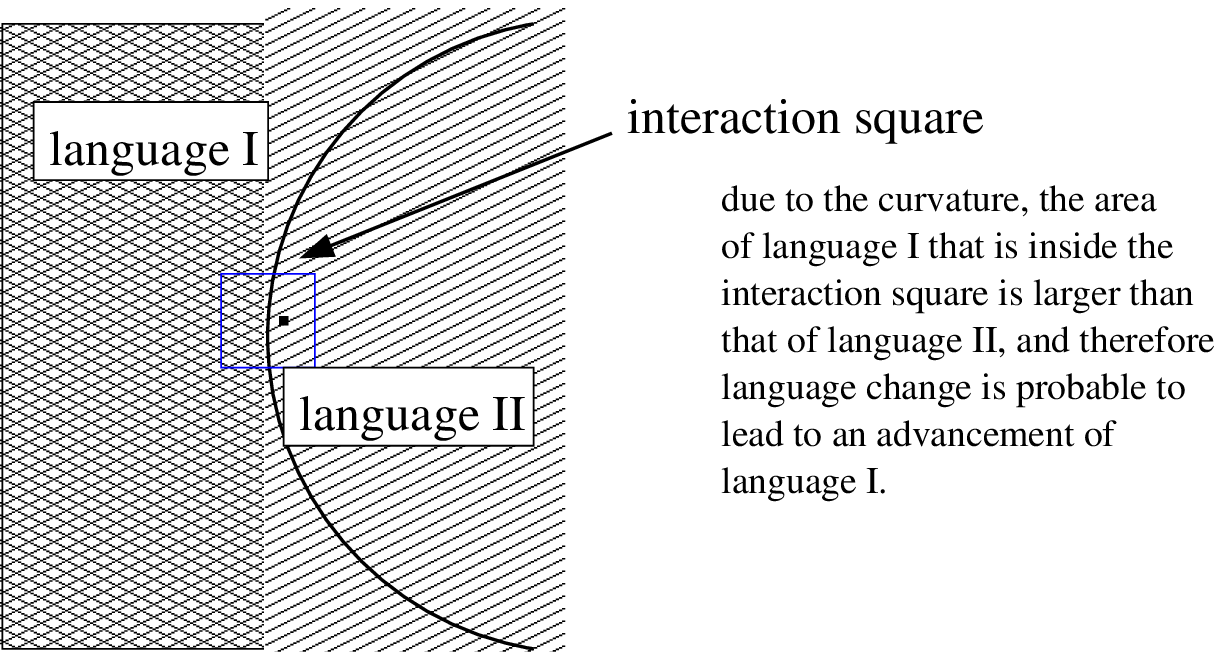} {Explanation
of instability of curved borders in the lattice simulation.}
%%%%%%%
that the interface between two languages is only stable if it is
straight (if $q > 0$): an individual placed on the border will
probably change its language to the one that is predominant within
the interaction range centered at it. A curvature in the interface
will make the population on the concave side of the curvature to
have less speakers within the interaction range than the
population on the other side; therefore, the former is disfavored
with respect to the latter language. Moreover, by this argument it
can be deduced that any cluster of one language completely
surrounded by sites with a different language will not survive;
due to the rule described above, it will shrink continuously,
until it becomes smaller than the interaction range; in this case
it quickly fades away as the ratio of the perimeter to the area of
the interaction range becomes larger and larger. This process can
be seen in simulation snapshots. In Figure \ref{fig:res:dying}
%%%%
\putfig{fig:res:dying}{13cm}{\figdir/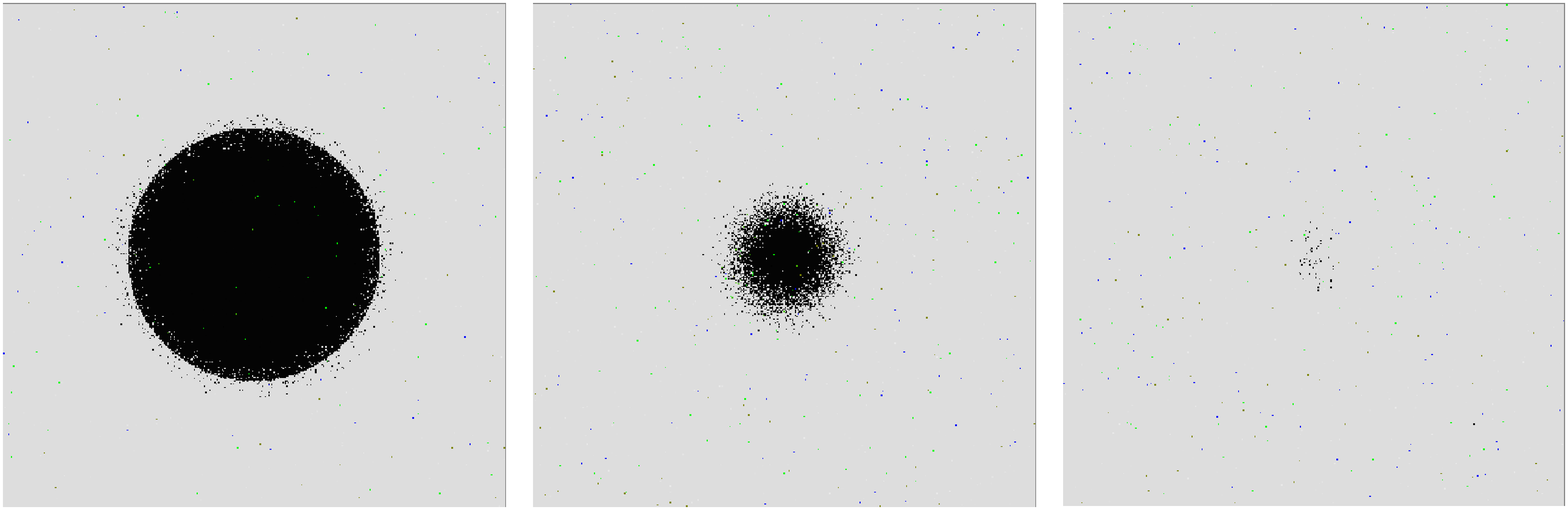}{Simulation of a language
surrounded by another language, in which the former disappears.}
%%%%
we see the time evolution (from left to right) of how a disk of
one language initially surrounded by a different language
eventually shrinks to zero. This fate of a language island would
change in an extended model, in which inhabitants of the island
can communicate (via internet, phone, satellite-TV) with remote
speakers of their own language.

In Figure \ref{fig:res:two}
%%%
\putfig{fig:res:two}{10cm} {\figdir/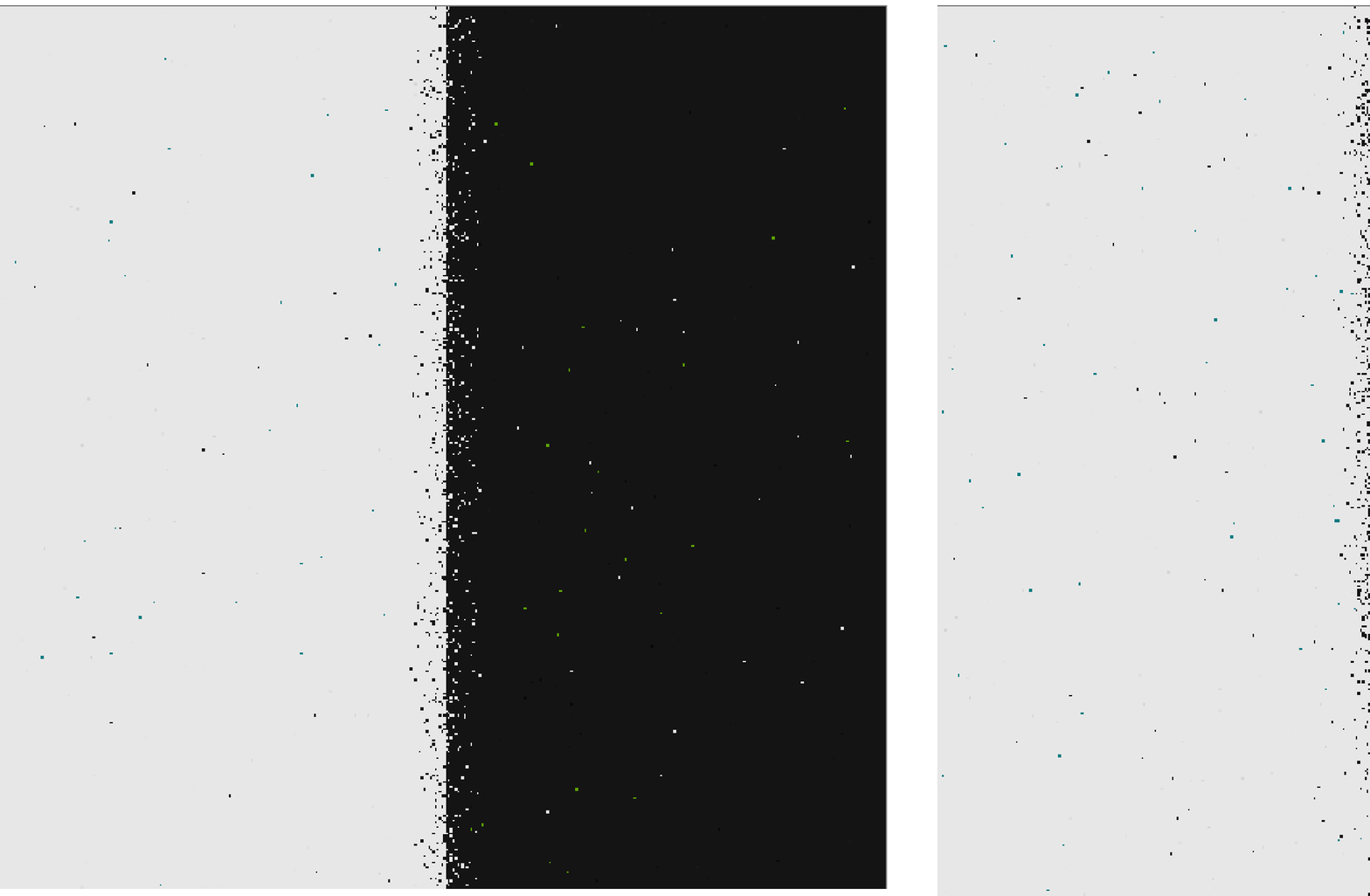}{Simulation of stable border
between two languages.}
%%%%%%%
we see a stable coexistence of two languages across a straight
border, with an inter-penetration depth that is determined by the
size of the interaction range. The figure on the left is taken soon
after the beginning of the simulation, while the one on the right
shows the stationary state that is finally reached. In an extended version of this model, one
could imitate different language policies in neighboring
countries, leading to asymmetrical penetration depths.

Beyond the coexistence of two languages along straight
borderlines, we simulated the coexistence of three languages,
Figure \ref{fig:res:three1} (again, time goes from left to right.)
The initial conditions had one language, in the shape of a square,
in the middle of a ``gradient'' of languages, where the languages
were the same along horizontal lines, and changing from line to
line. The gradient of languages started from language
$3500_{(16)}$ (again in hexadecimal notation) at the top of the
simulation space, and the language code (in terms of integers)
increased by one each line, up to $CB00_{(16)}$ at the bottom with
$n-16$ leading zero
bits. Due to this gradient, none of the
languages outside the square is strong enough, and the engulfed
language can ``capture'' the middle of the simulation lattice. The
gradient is transformed into clusters from which only the upper
and the lower two remain stable (as indicated in the outermost
right picture in Figure \ref{fig:res:three1}. Note that similar
shades of grey at the top and bottom of the picture do not
necessarily correspond to similar languages.)

%%%
\putfig{fig:res:three1}{13cm}{\figdir/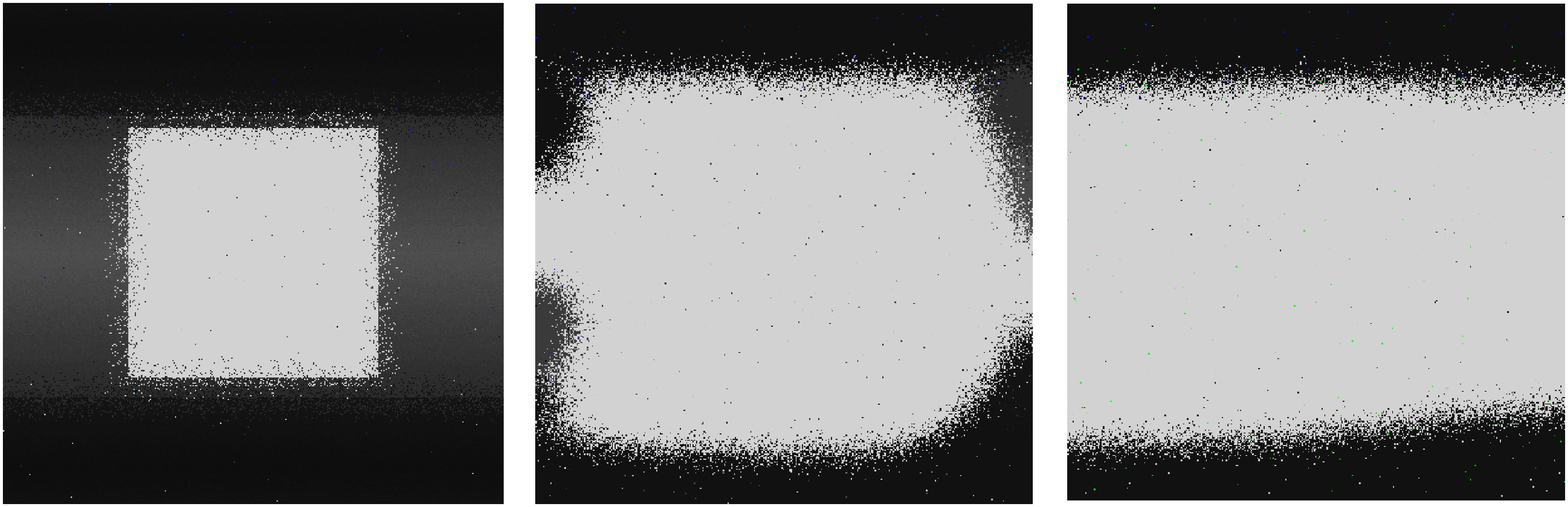}{Simulation of the
coexistence of three languages.}
%%%
\putfig{fig:res:three2}{10cm} {\figdir/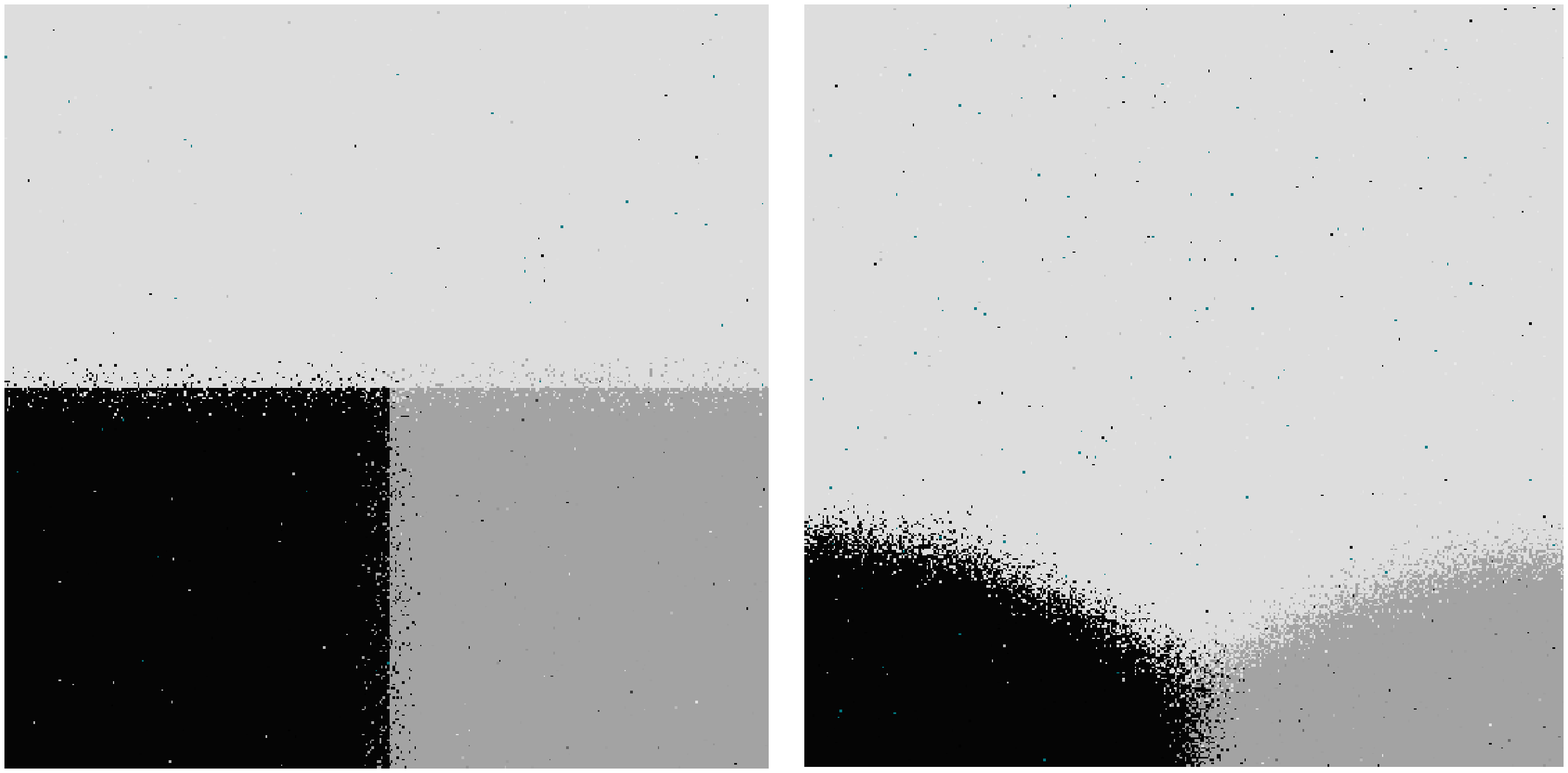}{Simulation
of three languages coexisting as a metastable state.}
%%%%%%%

The language evolution presented in Figure \ref{fig:res:three2}
(time from left to right) shows how a metastable state is reached,
where three languages survive. The state shown in the right
picture survives for some time, but ultimately the upper language
is able to completely eliminate the lower two ones, because it
initially occupies twice the area of the other languages, and due
to the interaction rules the probability for switching to another
language is proportional to the area this language occupies.

More interesting in this case is that the final state of the
simulation depends on the Hamming distance between the three
initial languages. In Figure \ref{fig:res:three2}, the Hamming
distance between any two of the initial languages was equal to
two. By changing only the bitstring describing the upper language,
so that the Hamming distance between it and the lower-left one was
equal to one, while the distance to the lower-right language
became three, we obtained the evolution from Figure
\ref{fig:res:two2}, where in the final state, two languages
survived instead of one.
%%%%%
\putfig{fig:res:two2}{10cm}{\figdir/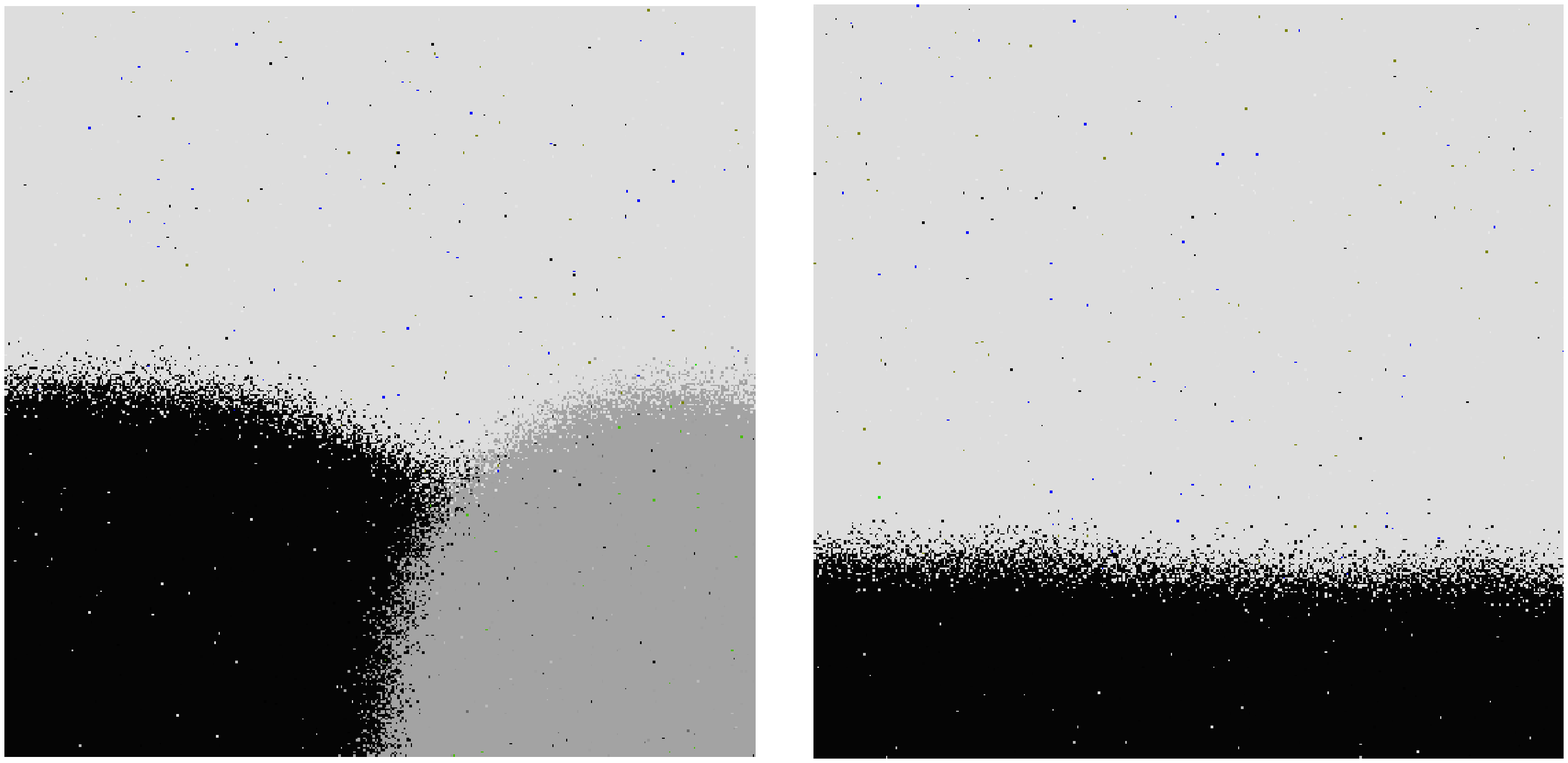}{Simulation
showing dependence of outcome on the Hamming distance between the
languages.}
%%%%%%%
If the language in the lower-left quarter of the plane is close in
terms of the Hamming distance to the language in the upper
half-plane, these two eliminate the third language and coexist as
a stable state. As explanation for the extinction of the third
language in Figure \ref{fig:res:two2}, we note that the upper and
lower-left coincide within the range of mutations if the Hamming
distance is small. An individual ``living'' at the confluence of
the three languages, is going to be biased toward changing its
language to the upper population, because that is predominant
within its interaction range. In the next time-step, only a
one-bit mutation can make this individual switch to the lower-left
language, while one time-step is not enough for it to mutate to
the lower-right one. In this way, the lower-left language has a
slight advantage with respect to the lower-right one, and we
obtain the observed outcome. Stated differently, the upper and
lower-left languages coincide within the typical size of
mutations; by arguments analogous to those of
Figure \ref{fig:res:dying}, the lower-right language then fades away.

It would be expected that even for larger values of the Hamming distances, for instance,
when the upper and the lower-left languages differ by 2 bits, while the difference between
the upper and the lower-right languages is 3 bits, the lower-left language would
still have a slight advantage, and would eventually overrun the lower-right one. That
is indeed observed in simulations.
As expected, however, the time needed to reach the final state is longer than in the previous case.

\section{Summary and Outlook}
\label{sec:outlook}
\subsection{Language population with random interactions}
%Within the model of Schulze and Stauffer, we studied a language
%population of individuals, each speaking one language described by a
%bitstring. They inherit their mother tongue from their parent with
%a one-bit mutation that happens with probability $p$, have
%offspring during their lifetime with probability $\alpha$,
%and die with probability $\propto \beta$. The parameter
%$\beta$ is chosen such that a stable population size $N_0$
%is achieved. During their lives, the individuals switch to
%another (randomly chosen) language with probability $\propto q$, thus favoring the
%language spoken by the majority. In our extended version of this model,
%individuals also favor languages that are close in Hamming distance to their
%own language.

In the original model of Schulze and Stauffer, we observe sharp
transitions between phases of language dominance and language
fragmentation, and along with this, transitions between small and
large Hamming distances between the two largest languages, both
types of transitions driven by the mutation parameter $p$ for
given language change parameter $q$ and total size $N$, or,
alternatively, driven by $q$ for given $p$ and $N$. The precise
location of the critical mutation rate depends on the language
change probability $q$, the initial conditions and the population
size. In the extended model ($m=4$), the transition between the
dominance and fragmentation phases is smoothed out, since the
switch to another language happens more conservatively; at the
same time, the transition in the Hamming distance is still sharp.
For $m=4$, the transitions are less dependent  on the initial
conditions, because dominance of a certain language is less
favored than in the original model.

In particular, we found in the scenario of language dominance that
a population evolution starting with one speaker initially, does
not alter the initial language if the population has reached its
stable size and prefers one dominant language. This is not
surprising although unrealistic, because the one-bit mutations
happen randomly and the switches favor dominance. For a stable
population size and two coexisting languages in the beginning, the
evolution favors a dominant language in the end that is one of the
coexisting initial languages rather than a ``superposition'' of
both.

\subsection{Language population on a square lattice}
The evolution of the language population on a square lattice was
visualized as a function of time. The square lattice was only used
to define the interaction range. A number of $m$ speakers was
randomly chosen out of the interaction range and the one with the
smallest Hamming distance to a given language was used in the
switch with probability $q$. Here we found a sensitive dependence
on the initial conditions, and depending on the mutual Hamming
distance, an initial set of two or three coexisting languages
turned out to give unstable, stable or metastable configurations.
Here further investigations are needed to predict the stability
behavior beyond the level of numerical evidence.

The models considered in this paper allow for several extensions.
%It is straightforward to study sexual reproduction, which could be
%related to the possibility of multilingual individuals. Another
%obvious generalization is to have different populations having
%different number of offspring per time-step, thus allowing for
%smaller languages to gain dominance if they have larger
%reproduction rates.
Biased mutations (with $r\geq 1$ bits mutated in a single event)
are a necessary condition for modelling language development from
its ancient to the modern version of the language. In the
so-called language change, we considered only a full replacement
of one language by another one, randomly chosen. Neither the
random one-bit mutations nor the full switches will ever mimic the
development of English out of ancient French and ancient German.
On a longer time-scale, the bitstring dynamics for languages
should imitate the interaction dynamics of parent genes in sexual
reproduction, resulting in the genes of their offspring. A
description of this dynamics would need, however, a deeper
understanding of language evolution from the viewpoint of
linguistics and a mapping between the bits and realistic traits or
features of languages.

\section{Acknowledgements}
It is a pleasure to thank D. Stauffer for drawing our attention to the topic of language competition.
%%%%%%%%%%%%%%%%%%%%%%%%%%%%%%%%%%%%%%%%%%%%%%%%%%%%%%%%%%%%%%%%%
%%%%%%%%%%%%%%%%%%%%%%%%%%%%%%%%%%%%%%%%%%%%%%%%%%%%%%%%%%%%%%%%%%%%%

%%%%%%%%%%%%%%%%%%%%%%%%%%%%%%%%%%%%%%%%%%%%%%%%%%%%%%%%%%%%%%%%%%%%%
\end{document}